\documentclass[aps,pre,twocolumn,showpacs,superscriptaddress]{revtex4}
\usepackage{graphicx}

\begin{document}

\title{Quasispecies dynamics on a network of interacting genotypes and
idiotypes: Formulation of the model}

\author{Valmir C. Barbosa}
\affiliation{Programa de Engenharia de Sistemas e Computa\c c\~ao, COPPE,
Universidade Federal do Rio de Janeiro,
Caixa Postal 68511, 21941-972 Rio de Janeiro - RJ, Brazil}

\author{Raul Donangelo}
\affiliation{Instituto de F\'\i sica,
Universidade Federal do Rio de Janeiro,
Caixa Postal 68528, 21941-972 Rio de Janeiro - RJ, Brazil}
\affiliation{Instituto de F\'\i sica, Facultad de Ingenier\'\i a,
Universidad de la Rep\'ublica,
Julio Herrera y Reissig 565, 11.300 Montevideo, Uruguay}

\author{Sergio R. Souza}
\affiliation{Instituto de F\'\i sica,
Universidade Federal do Rio de Janeiro,
Caixa Postal 68528, 21941-972 Rio de Janeiro - RJ, Brazil}
\affiliation{Instituto de F\'\i sica,
Universidade Federal do Rio Grande do Sul,
Caixa Postal 15051, 91501-970 Porto Alegre - RS, Brazil}

\begin{abstract}
A quasispecies is the stationary state of a set of interrelated genotypes that
evolve according to the usual principles of selection and mutation. Quasispecies
studies have invariably concentrated on the possibility of errors during
genotype replication and their role in promoting either the survival or the
demise of the quasispecies. In a previous work [V. C. Barbosa, R. Donangelo, and
S. R.  Souza, J. Theor. Biol.\ \textbf{312}, 114 (2012)], we introduced a
network model of quasispecies dynamics, based on a single probability parameter
($p$) and capable of addressing several plausibility issues of previous models.
Here we extend that model by pairing its network with another one aimed at
modeling the dynamics of the immune system when confronted with the
quasispecies. The new network is based on the idiotypic-network model of
immunity and, together with the previous one, constitutes a network model of
interacting genotypes and idiotypes. The resulting model requires further
parameters and as a consequence leads to a vast phase space. We have focused on
a particular niche in which it is possible to observe the trade-offs involved in
the quasispecies' survival or destruction. Within this niche, we give simulation
results that highlight some key preconditions for quasispecies survival. These
include a minimum initial abundance of genotypes relative to that of the
idiotypes and a minimum value of $p$. The latter, in particular, is to be
contrasted with the stand-alone quasispecies network of our previous work, in
which arbitrarily low values of $p$ constitute a guarantee of quasispecies
survival.
\end{abstract}


\maketitle

\section{Introduction}
\label{sec:intro}

The immune system is one of the central regulatory systems of the body, being
responsible for the detection and eventual removal of both external agents that
may be potentially harmful and body cells whose behavior may have become
abnormal. These targets of the immune system are generically referred to as
antigens, in allusion to the fact that they trigger the system's response and
ultimate production of antibodies or activation of destroying cells. In order to
be able to respond appropriately to a potentially large variety of antigens, the
immune system is highly adaptive and along its existence evolves from an initial
state of innate immunity through a series of states of acquired immunity.
Modeling the dynamics that gives rise to this type of learning has been
challenging for many decades and has elicited the appearance of at least two
main explanatory frameworks. One of them is the clonal-selection theory
\cite{b59,f95}, which postulates the preferential survival of those cell types
that are more effective for a certain class of antigens. This theory has been
successful in many respects but has failed in others, e.g., explaining the
existence of any level of innate immunity, which by definition exists in the
absence of any antigens. 

The other leading framework to model the dynamics of immunity is that of the
so-called idiotypic network \cite{j74}. The main idea is that the molecular
structures capable of being recognized by the immune system, known as epitopes,
are found not only in antigens but also in the receptors of the immune-system
cells whose task is to recognize those antigens. That is, not only can these
cells recognize antigens, they can also recognize one another in much the same
way. Readily, the existence of such a network of epitope types, known as
idiotypes, has the potential to explain not only how immunity is acquired but
also how it can exist before antigens are ever encountered. The
idiotypic-network theory has enjoyed both enthusiasm and skepticism along the
years, the latter owing mainly to the many difficulties associated with
confirming it experimentally. Many of its elements, however, are present in
models of a hybrid nature, particularly in those that aim to characterize those
phenomena, such as autoimmunity, that are essentially of a systemic nature
(cf., e.g., \cite{fabc04,b07,mkbmqtcb11} and references therein).

One common element of all mathematical models of the immune system is the
overly simplistic manner in which the system's interaction with antigens is
handled. Typically the immune-system model involves a set of time-dependent
equations describing the behavior of certain quantities (e.g., the abundance of
a given cell type) and including independent terms to account for the presence
of antigens. Such terms can be adjusted to account for the simultaneous presence
of several antigens, but in general independently of one another. The drawback,
of course, is that in important cases such as infections by some viral species,
the many different viral strains that may be concomitantly present mutate into
one another frequently and confront the immune system in ever-changing ways.

Here we address the problem of how antigens and entities of the immune system 
interact with one another when both sides already display complex dynamic
behavior even when left to themselves. We introduce a random-graph model of this
interaction that takes into account not only the interaction itself but also the
dynamics of genotype mutation on the antigen side and of idiotype recognition on
the immune-system side. Our model can be viewed as comprising two subnetworks,
one whose nodes represent genotypes that mutate into one another as they
replicate, the other whose nodes represent idiotypes that stimulate (are
recognized by) one another and proliferate as a consequence. The two subnetworks
are put together by the addition of new edges to give idiotypes the further
stimuli provided by the genotypes.

In building the two subnetworks we have drawn on previous ideas dealing with
graph-theoretic representations of both interacting genotypes and idiotypes. On
the genotype side the network we use is the same we introduced recently
\cite{bds12} to model the dynamics of the so-called quasispecies. A quasispecies
is the stationary state of a set of genotypes that mutate into one another while
replicating without recombination based on fitnesses that do not depend on
genotype abundance (cf., e.g., \cite{e71,es77,d09,la10,mlcgm10} and references
therein). Our model is based on a single probability parameter, $p$, that
regulates both graph connectivity and the occurrence of mutations, and can also
successfully account for the well-known transition from adaptation to degeneracy
when mutations become too frequent. For the idiotype side we use a similar
network while incorporating some of the essential premises of the
idiotypic-network theory. Like our quasispecies network, our idiotypic network
is based on a single probability parameter, $r$, whose function is to regulate
both graph connectivity and the occurrence of stimulation/recognition of
idiotypes by one another.

Each node in our model is a binary sequence, i.e., a sequence of $0$'s and
$1$'s. This representation is suitable for both genotypes and idiotypes, since
it can easily accommodate the types of interaction that are necessary in each
subnetwork. Specifically, on the quasispecies side mutations are bit alterations
and are more likely to occur between two genotypes that do not differ at too
many loci. On the idiotypic side, by contrast, stimulation occurs as a function
of how well the two sequences involved can be said to be complementary to each
other on a locus-wise basis, being more likely as complementarity increases.
This stems directly from the physicochemical nature of the molecular coupling
that, in the immune system, leads to stimulation. As the quasispecies and
idiotypic networks are combined together into the full model, this same notion
of complementarity is used to account for how genotypes stimulate idiotypes,
again based on probability $r$.

Before proceeding, we note that, by virtue of its underlying random-graph model
and of the agent interactions this model represents, the subject we study in the
present work is related to various other topics of interest. One of them is the
wider discipline of complex networks, which in little more than a decade has
uncovered many common underpinnings to a great variety of natural and
technological systems of interacting agents \cite{bs03,nbw06,bkm09}. Another of
these topics is that of evolutionary games, which encompass the so-called
predator-prey systems and, essentially, are characterized by the mediation of
abundance-dependent fitnesses. The reader is referred to \cite{n06,kkwf13}, for
example, as well as to references therein.

In a similar vein, note that the essence of our model is that it portrays the
interaction of antagonistic agent populations as they adapt to each other while
competing for supremacy. Thus, while in the present study we instantiate such
agents by interrelated genotypes invading the body and idiotypes defending it,
taking a broader view may come to benefit several other areas. One area with
obvious conceptual ties to our invader-defender setting is that of computer
security, and in fact proposals have been put forward that are based on
immune-related notions \cite{hwgl02,dss11}. This has also been the case of other
areas with less obvious but equally relevant connections to our study, such as
combinatorial optimization (cf.\ \cite{cnpt07} for an application to the
prediction of protein structure), user profiling for adaptive information
filtering in online systems \cite{nvr10}, and classification of textual
documents \cite{ar11}. Useful reviews on how these and other areas have come to
be influenced by immune-inspired abstractions can be found in \cite{g05,thsc08}.

The remainder of the paper is organized in the following manner. First we
introduce the model's details in Section~\ref{sec:model}. Then we discuss some
aspects of the computational methodology we have used and present our results,
all in Section~\ref{sec:results}. These results elucidate the role played by
initial conditions, as well as by one of our main parameters, the probability
$p$, in promoting the survival or the demise of the quasispecies as a function
of how genotypes interact with one another and with the idiotypes. These issues
are discussed in Section~\ref{sec:disc}, which is followed by conclusions in
Section~\ref{sec:concl}.

\section{Model}
\label{sec:model}

We consider $2^{L+1}$ binary sequences of length $L$ and group them into sets
$A$ and $B$, each set comprising all $2^L$ distinct sequences of length $L$.
Each sequence in set $A$ stands for a genotype, each sequence in set $B$ for an
idiotype. Our model is based on a directed random graph $D$ of node set
$A\cup B$. For $i\in A\cup B$, we use $I_i$ to denote the set of in-neighbors of
node $i$ and $O_i$ to denote its set of out-neighbors. Graph $D$ has self-loops
at all nodes, so it holds that both $i\in I_i$ and $i\in O_i$. We describe the
edge set of graph $D$ in three separate stages: first the edges involving nodes
in $A$ exclusively, then edges involving nodes in $B$ exclusively, then the
edges used to interconnect the two halves.

\subsection{Quasispecies network}

For $i,j\in A$ (i.e., $i$ and $j$ are both genotypes), possibly identical, an
edge exists directed from $i$ to $j$ with probability $p_{ij}=p^{H_{ij}}$, where
$p$ is a probability parameter and $H_{ij}$ is the Hamming distance between $i$
and $j$. If the edge does exist then it is possible for genotype $i$ to mutate
into genotype $j$ during replication, provided $i\neq j$. The probability that
such mutation occurs is $q_{ij}$, assumed proportional to $p_{ij}$ in such a way
that $\sum_{j\in O_i}q_{ij}=1$, where $q_{ii}$ is the probability that genotype
$i$ remains unchanged during replication. Note that larger Hamming distances
entail smaller connection probabilities and, consequently, smaller mutation
probabilities as well.

Letting $X_i$ denote the time-dependent abundance of genotype $i$, we write
\begin{equation}
\dot{X}_i=\sum_{j\in I_i}f_jq_{ji}X_j,
\label{eq:XA1}
\end{equation}
where $f_j$ is the fitness of genotype $j$ and reflects its replication rate. We
assume throughout that $f_j=2^{-d_j}$, where $d_j$ is the number of $1$'s in
genotype $j$. That is, a genotype's fitness decays exponentially with its
Hamming distance to the genotype having no $1$'s (the fittest genotype, or wild
type). Equation~(\ref{eq:XA1}) is the well-known quasispecies equation
\cite{be06,n06}, now written for graph $D$ as in the uniform case of
\cite{bds12}.

\subsection{Idiotypic network}

For $i,j\in B$ (i.e., $i$ and $j$ are both idiotypes), again possibly identical,
an edge exists directed from $i$ to $j$ with probability $r_{ij}=r^{L-H_{ij}}$,
where $r$ is another probability parameter. The existence of this edge indicates
that idiotype $i$ stimulates idiotype $j$ during the proliferation phase of the
idiotypic dynamics, provided $i\neq j$. The probability that this stimulation
occurs is $s_{ij}$, assumed proportional to $r_{ij}$ in such a way that
$\sum_{j\in O_i}s_{ij}=1$, where $s_{ii}$ is the probability that idiotype $i$
stimulates no other idiotype during proliferation. Now larger Hamming distances
entail larger connection and stimulation probabilities, which indeed are
expected to be larger if the idiotypes involved are more complementary to each
other (i.e., differ at more loci).

Letting $X_i$ denote the time-dependent abundance of idiotype $i$, and assuming
that $X_i$ grows in proportion by some rate $\lambda>0$ to the total stimulus
received by $i$, yields
\begin{equation}
\dot{X}_i=\lambda\sum_{j\in I_i}s_{ji}X_j.
\label{eq:XB}
\end{equation}
This constitutes a very simple model of the idiotypic network in the absence of
antigens.

\subsection{Genotype-idiotype interaction network}

We are now in position to describe the entirety of graph $D$, of node set
$A\cup B$, which is intended to model the interaction of the quasispecies and
idiotypic networks, of node sets $A$ and $B$, respectively. We do this by
allowing the existence of edges directed from nodes in $A$ to nodes in $B$,
indicating that idiotypes are stimulated not only by one another but also by
the genotypes against which they are supposed to constitute a defense. This, in
effect, brings antigens into the idiotypic dynamics and lets graph $D$ work as a
model of how invading genotypes and defending idiotypes interact with one
another. Adding these edges to the network has the potential of enlarging the
$O_i$ sets for $i\in A$ and the $I_i$ sets for $i\in B$. The former has the
immediate consequence of requiring that we adjust the quasispecies-network
constraint $\sum_{j\in O_i}q_{ij}=1$ to $\sum_{j\in O_i\cap A}q_{ij}=1$. As for
the latter, it remains for us to specify how a genotype $i\in A$ stimulates an
idiotype $j\in B$.

We begin by assuming that the length-$L$ sequences representing genotypes and
idiotypes are all relative to the same support, i.e., they all share the same
representation space, $\{0,1\}^L$. In other words, the Hamming distance is well
defined also between a genotype and an idiotype. Our approach is then to handle
idiotype stimulation independently of whether it is effected by a genotype or
another idiotype. That is, we let the edge from genotype $i\in A$ to idiotype
$j\in B$ exist with the same probability $r_{ij}$ as above. When it does exist,
stimulation too occurs with the same probability $s_{ij}$ as above, so every
node $i\in A$ is now subject to the additional constraint that
$\sum_{j\in O_i\cap B}s_{ij}=1$. Equation~(\ref{eq:XB}), therefore, remains
unchanged.

Once coupled in this way, the $2^{L+1}$ differential equations given in
Eqs.~(\ref{eq:XA1}) and~(\ref{eq:XB}) mandate an unbounded exponential growth of
both genotype and idiotype abundances from any nontrivial initial values. We
therefore lack further coupling in order for the possibility of genotype removal
by the idiotypes to be explicitly taken into account. We achieve this by
rewriting Eq.~(\ref{eq:XA1}) as
\begin{equation}
\dot{X}_i=\sum_{j\in I_i}f_jq_{ji}X_j-\mu\sum_{j\in O_i\cap B}s_{ij}X_j,
\label{eq:XA2}
\end{equation}
where $\mu>0$ is a rate parameter. This modification to Eq.~(\ref{eq:XA1}) lets
the abundance of genotype $i$ be decreased at a rate that is proportional to how
strongly $i$ stimulates each idiotype $j$.

Further modifications to all equations might still be considered for the removal
of genotypes as they mutate into other genotypes or the removal of idiotypes as
they stimulate (and are therefore recognized and sought for destruction by)
other idiotypes. We approach this last step by henceforth considering relative,
rather than absolute, abundances. Proceeding in this way is equivalent to
imposing that the various abundances sum up to a constant at all times, which
automatically leads to the appearance of terms that account for the desired
sources of both genotype and idiotype removal.

For $i\in A\cup B$, let $x_i=X_i/\sum_{k\in A\cup B}X_k$ be the relative
abundance of genotype or idiotype $i$. Thus, $\sum_{i\in A\cup B}x_i=1$ at all
times. It follows that
\begin{eqnarray}
\dot{x}_i
&=&\frac{\dot{X}_i}{\sum_{k\in A\cup B}X_k}-
x_i\frac{\sum_{k\in A\cup B}\dot{X}_k}{\sum_{k\in A\cup B}X_k}\cr
&=&\frac{\dot{X}_i}{\sum_{k\in A\cup B}X_k}-
x_i(\phi-\mu\psi+\lambda),
\label{eq:x*}
\end{eqnarray}
where $\phi=\sum_{k\in A}f_kx_k$ and
$\psi=\sum_{k\in B}x_k\sum_{\ell\in I_k\cap A}s_{\ell k}$. In this expression
for $\psi$, the rightmost summation represents the stimulatory influence of all
genotypes upon idiotype $k$ and is here referred to as that idiotype's
proliferability. Note, additionally, that $\phi/\sum_{k\in A}x_k$ and
$\psi/\sum_{k\in B}x_k$ are, respectively, the average genotype fitness and
idiotype proliferability.

Equation~(\ref{eq:x*}) yields the final expressions for $\dot{x}_i$:
\begin{equation}
\dot{x}_i
=\sum_{j\in I_i}f_jq_{ji}x_j
-\mu\sum_{j\in O_i\cap B}s_{ij}x_j
-x_i(\phi-\mu\psi+\lambda)
\label{eq:xA}
\end{equation}
for $i\in A$ and
\begin{equation}
\dot{x}_i
=\lambda\sum_{j\in I_i}s_{ji}x_j
-x_i(\phi-\mu\psi+\lambda)
\label{eq:xB}
\end{equation}
for $i\in B$.

\subsection{Expected connectivity}

In graph $D$, a genotype $i$ may have out-neighbors both in set $A$ (other
genotypes) and in set $B$ (idiotypes). If one ignores all self-loops, then the
expected number of out-neighbors of the former type can be obtained by
considering every other genotype $j$ and counting the edge from $i$ to $j$ with
weight $p^{H_{ij}}$ \cite{k90}. That is, a genotype's expected number of
out-neighbors that are genotypes other than itself is
\begin{equation}
\sum_{h=1}^L{L\choose h}p^h=(1+p)^L-1.
\label{eq:outAA}
\end{equation}
As for a genotype's expected number of out-neighbors that are idiotypes, we
have
\begin{equation}
\sum_{h=0}^L{L\choose h}r^{L-h}=(1+r)^L.
\label{eq:outAB}
\end{equation}

An idiotype, on the other hand, can only have other idiotypes as out-neighbors.
Its number of out-neighbors that are also idiotypes, ignoring self-loops, is
\begin{equation}
\sum_{h=1}^L{L\choose h}r^{L-h}=(1+r)^L-r^L.
\label{eq:outBB}
\end{equation}

\section{Results}
\label{sec:results}

We begin by revisiting Eq.~(\ref{eq:xA}) and recognizing that it is possible for
$\dot{x}_i$ to be negative when $x_i=0$, thus leading the implicit constraint
that $x_i\ge 0$ for all $i\in A\cup B$ at all times to be violated. We fix this
by noting that it suffices that this equation's predecessor for absolute
abundances, Eq.~(\ref{eq:XA2}), be modified to
\begin{equation}
\dot{X}_i=\sum_{j\in I_i}f_jq_{ji}X_j-\mu H(X_i)\sum_{j\in O_i\cap B}s_{ij}X_j,
\end{equation}
where $H(z)$ is the Heaviside step function, adapted to yield $1$ if and only
if $z>0$ (yield $0$, otherwise). This modification forbids $\dot{X}_i$ to
be negative when $X_i=0$ and affects both Eqs.~(\ref{eq:xA}) and~(\ref{eq:xB}),
since the expression for $\psi$ is itself affected.

The actual expressions we use are then as follows, where all appearances of the
$H$ function have been shifted by a suitably small $\delta>0$ to avoid numerical
instabilities. For
\begin{equation}
\psi=\sum_{k\in B}x_k\sum_{\ell\in I_k\cap A}s_{\ell k}H(x_\ell-\delta),
\label{eq:psif}
\end{equation}
we have
\begin{equation}
\dot{x}_i
=\sum_{j\in I_i}f_jq_{ji}x_j
-\mu H(x_i-\delta)\sum_{j\in O_i\cap B}s_{ij}x_j
-x_i(\phi-\mu\psi+\lambda)
\label{eq:xAf}
\end{equation}
for $i\in A$ and
\begin{equation}
\dot{x}_i
=\lambda\sum_{j\in I_i}s_{ji}x_j
-x_i(\phi-\mu\psi+\lambda)
\label{eq:xBf}
\end{equation}
for $i\in B$.

While Eqs.~(\ref{eq:xAf}) and~(\ref{eq:xBf}) provide instantaneous values of
$\dot{x}_i$ that are correct in the sense of never allowing $x_i$ to fall below
$\delta$, actually solving the equations requires further control. Specifically,
when the value of $x_i$ is to be updated, say at time $t$, one must ensure that
the time step to be used, $\Delta t$, necessarily entails
$x_i+\dot{x}_i\Delta t\ge\delta$. Thus, before stepping time when $\dot{x}_i<0$
one must first discover the time $T>t$ at which this derivative will lead to
$x_i=\delta$ [that is, one must solve $x_i+\dot{x}_i(T-t)=\delta$ for $T$] and
then choose $\Delta t<T-t$. What we do is solve $x_i+\dot{x}_i(T-t)=\delta/10$
instead (again, to avoid numerical instabilities), which implies that
$x_i>\delta/10$ at all times. Consequently, even though Eqs.~(\ref{eq:xAf})
and~(\ref{eq:xBf}) ensure $\dot{x}_i\ge 0$ already for $x_i=\delta$, during
solution it is still possible for $x_i$ to fall below $\delta$, though remaining
strictly above $\delta/10$. We use $\delta=10^{-10}$ throughout.

We give results for $L=10$ (i.e., for $1\,024$ genotypes and $1\,024$ idiotypes)
in Figs.~\ref{fig:fig1}--\ref{fig:fig6}. These results reflect our exploration
of a specific parameter niche in which it is possible to observe a rich variety
of behaviors for the entire quasispecies
(Figs.~\ref{fig:fig1}--\ref{fig:fig3}) and, particularly, for the wild type
(Figs.~\ref{fig:fig4}--\ref{fig:fig6}). We follow \cite{bds12} and let $x_1$
denote the relative abundance of the wild type. Moreover, we let the relative
abundance of genotypes be denoted by $x_A$, i.e.,
\begin{equation}
x_A=\sum_{i\in A}x_i.
\end{equation}
We use $x_A(0)$ to denote the initial value of $x_A$.

All figures give results based on at least $10^4$ instances of graph $D$.
Solving Eqs.~(\ref{eq:xAf}) and~(\ref{eq:xBf}) for each instance starts at
uniform relative abundances for genotypes and likewise for idiotypes. That is,
initially $x_i=x_A(0)/2^L$ for $i\in A$ and $x_i=[1-x_A(0)]/2^L$ for $i\in B$.
All $2^{L+1}$ equations are then time-stepped through $t=50$, which empirically
we found to suffice for all relative abundances to reach a stationary state.

\section{Discussion}
\label{sec:disc}

\begin{figure}
\includegraphics[scale=0.41]{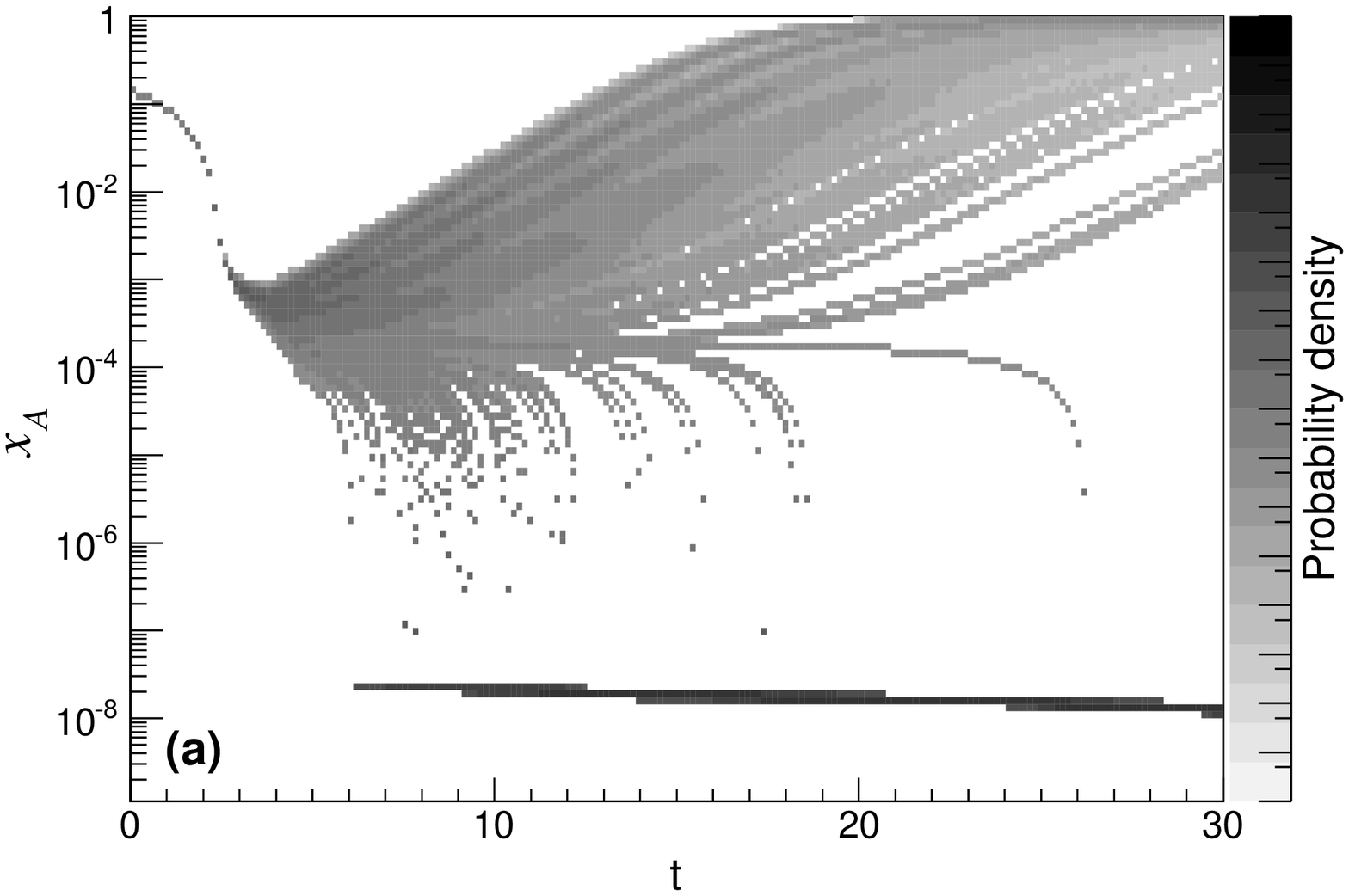}
\includegraphics[scale=0.41]{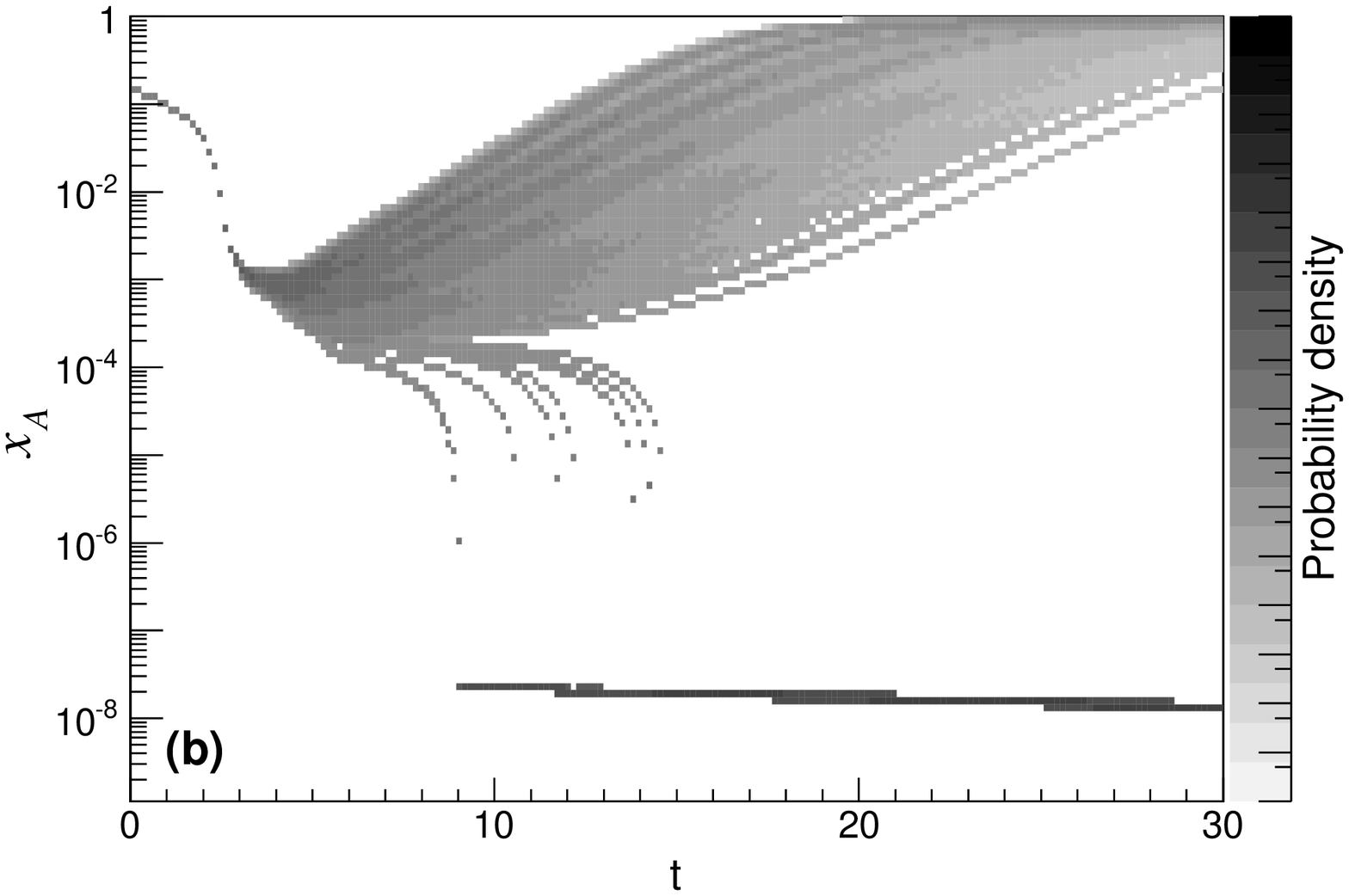}
\includegraphics[scale=0.41]{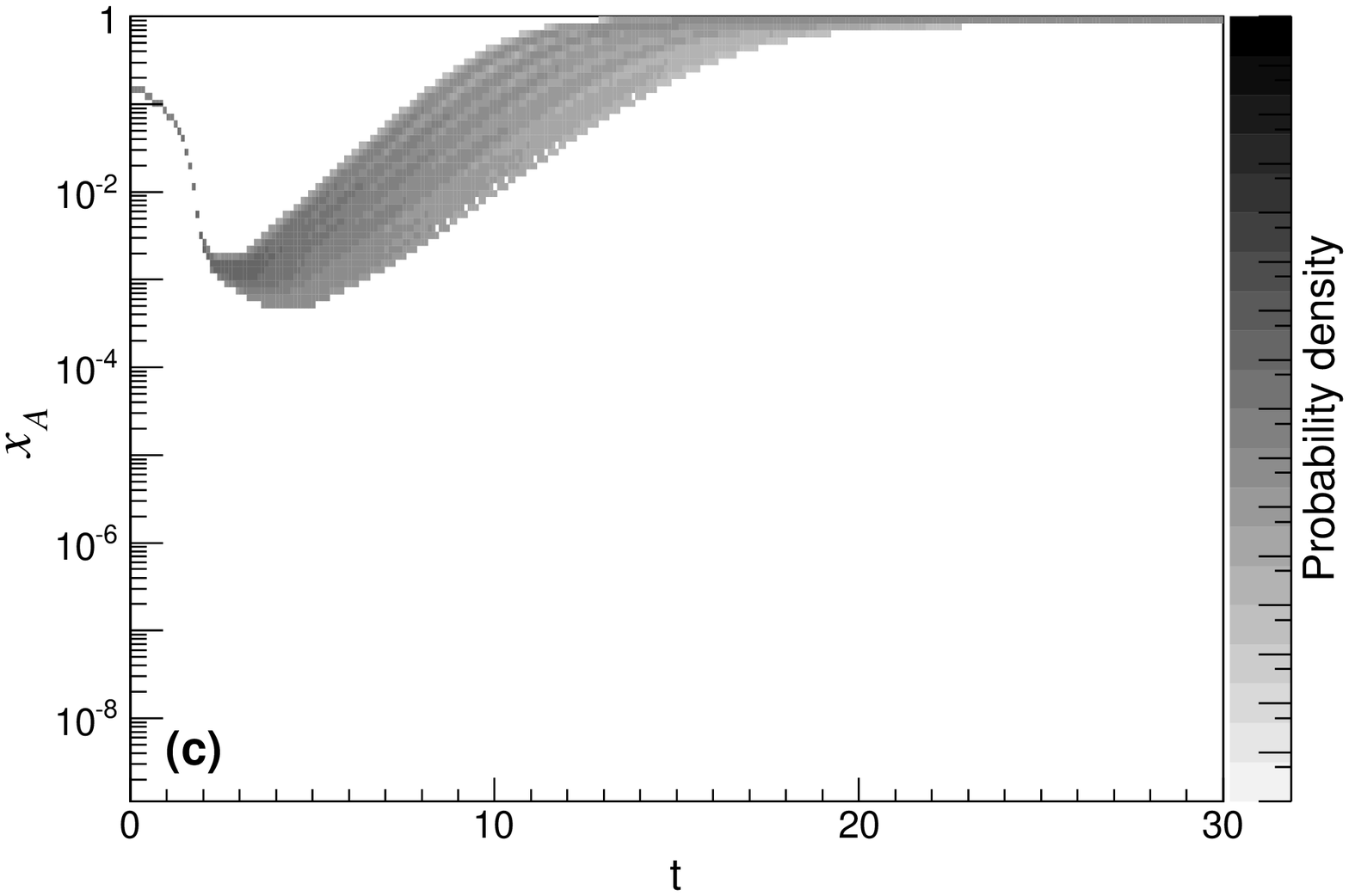}
\caption{Evolution of the probability distribution of $x_A$ from $x_A(0)=0.14$
(a), $x_A(0)=0.15$ (b), and $x_A(0)=0.16$ (c), for $p=r=0.1$ and
$\lambda=\mu=0.1$. Data are based on at least $10^4$ independent instances of
graph $D$ and are log-binned to the base $1.2$. Probabilities are given
according to the gray scale on the right of each panel, where darker is closer
to $1$.}
\label{fig:fig1}
\end{figure}

The three plots in Fig.~\ref{fig:fig1} show the distribution of the genotypes'
relative abundance, $x_A$, as $t$ is varied from $t=0$ through $t=30$. All cases
correspond to $p=r=0.1$, meaning that on average both the genotype and the
idiotype subnetworks have densities of the same order of magnitude, and so does
the set of edges directed from the genotype subnetwork to the idiotype
subnetwork to account for the removal of genotypes by idiotypes
[cf.\ Eqs.~(\ref{eq:outAA})--(\ref{eq:outBB})]. It also means that the dynamics
of genotype mutation and of idiotype stimulation are based on the same
underlying probability. The three cases also have in common that
$\lambda=\mu=0.1$, so the rate at which idiotypes proliferate due to stimulation
by other idiotypes and the rate at which genotypes are eliminated as they
stimulate idiotypes are the same. What distinguishes the three cases from one
another is the initial relative abundance of genotypes, $x_A(0)$, which is
varied from $0.14$ in panel (a) through $0.16$ in panel (c).

These three values of $x_A(0)$ were singled out, despite being so close to one
another, because they allow us to qualitatively zoom in on what appears to be a
transition from a regime in which the genotypes may either disappear or endure
in the long run (i.e., the quasispecies may perish or survive) to another in
which they endure almost certainly. In fact, examining the three panels of
Fig.~\ref{fig:fig1} reveals that, even though in all three cases the genotypes
are driven toward a sharp decrease in relative abundance up to about $t=2$,
regardless of the particular instance of graph $D$ on which the dynamics is
taking place, thereafter network structure begins to matter and does so in a
manner that depends on what the relative genotype abundance was to begin with.

For the lowest of the three values used in the figure [$x_A(0)=0.14$], and
notwithstanding the fact that there exist network topologies continuing to push
the genotypes toward the demise of the quasispecies, there are also cases in
which the underlying network topology supports the survival of the quasispecies
to the point that it seriously threatens the idiotype population, i.e., $x_A$
approaches $1$. This turn of events becomes significantly more pronounced as
$x_A(0)$ is slightly increased, quickly reaching the situation in which almost
no network topology supports the destruction of the quasispecies [for
$x_A(0)=0.16$].

\begin{figure}
\includegraphics[scale=0.41]{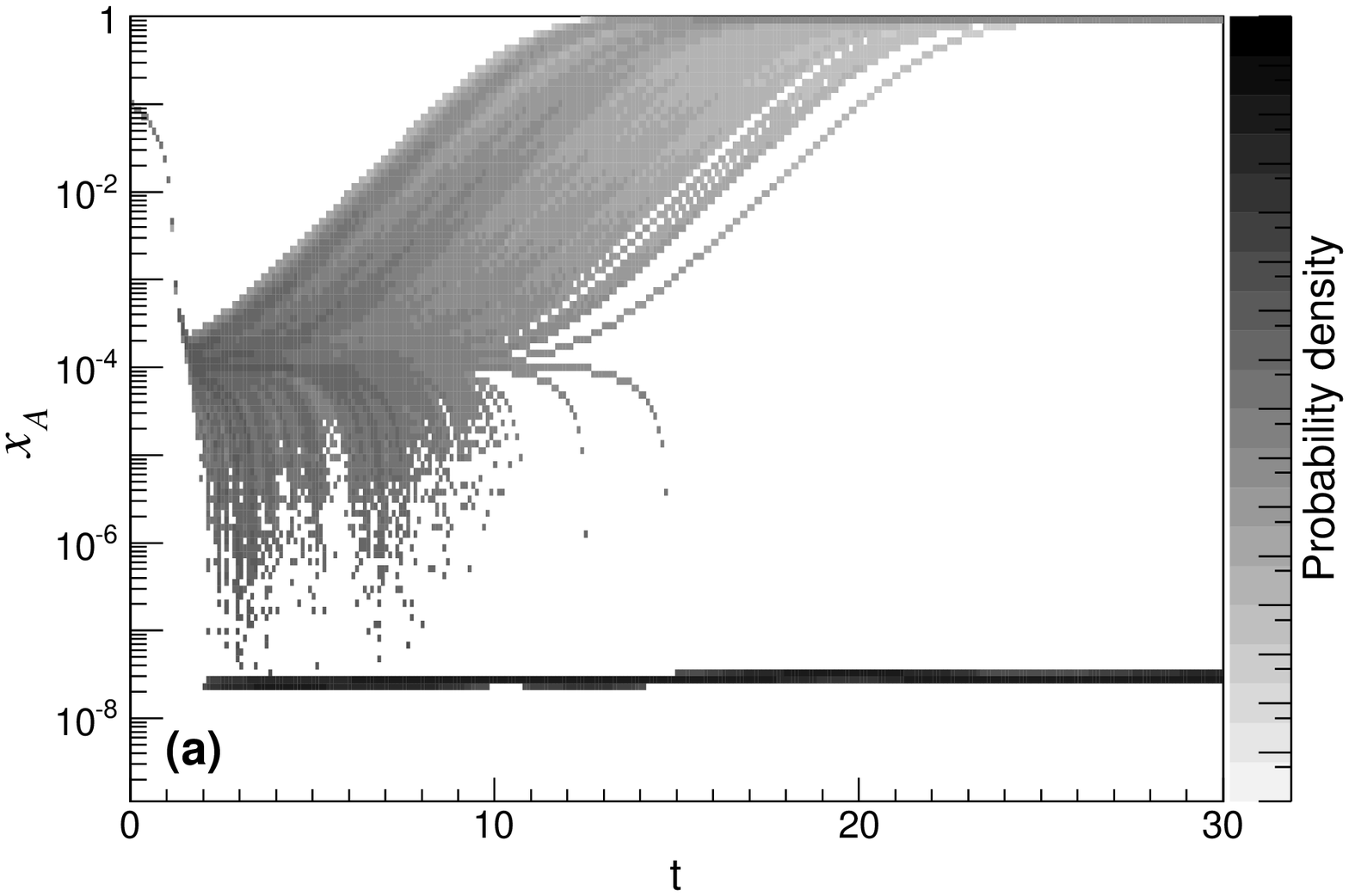}
\includegraphics[scale=0.41]{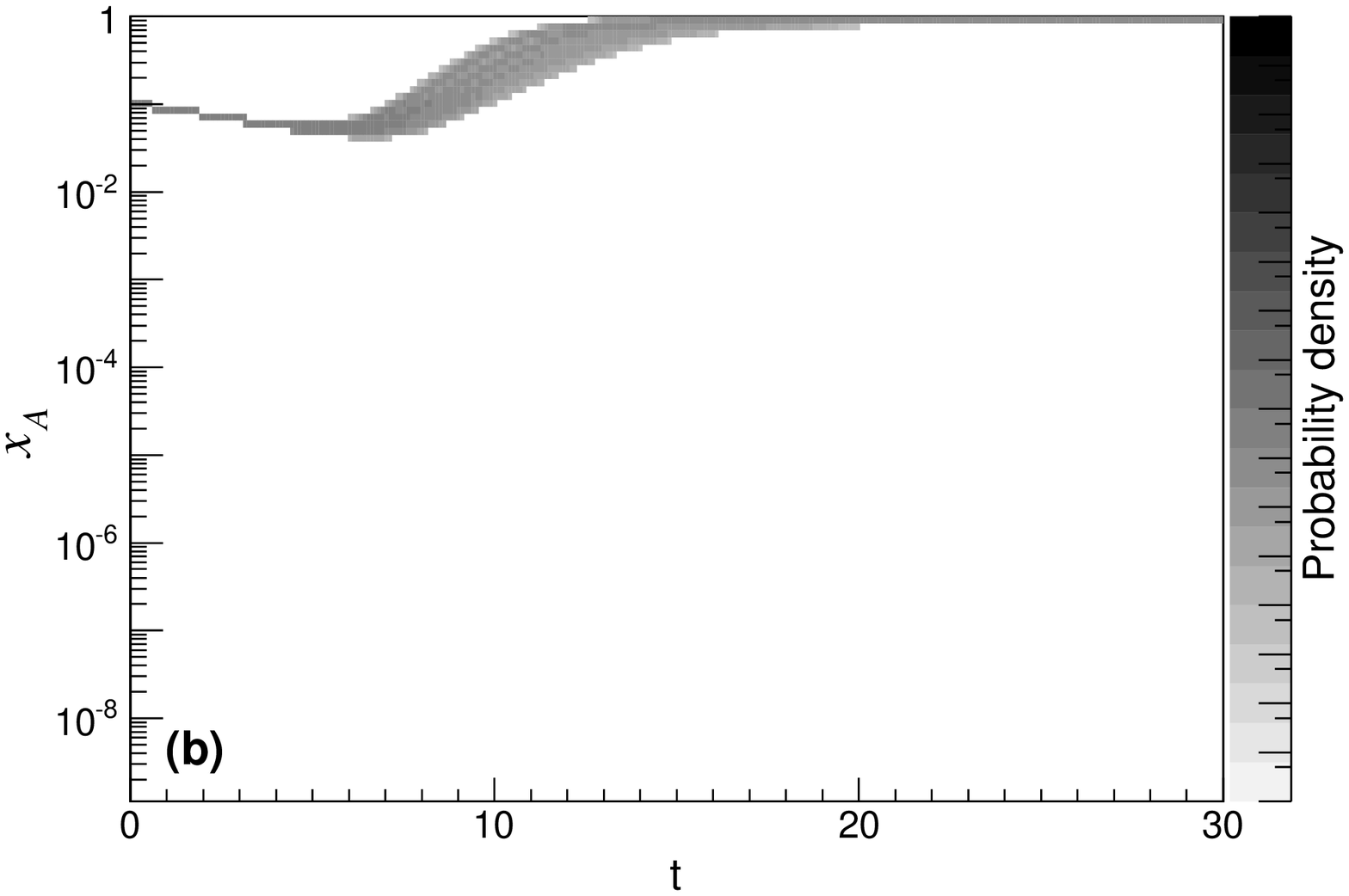}
\caption{Evolution of the probability distribution of $x_A$ from $x_A(0)=0.10$
for $p=r=0.1$. Panel (a) has $\lambda=0.01$ with $\mu=0.1$; panel (b) has
$\lambda=0.1$ with $\mu=0.01$. Data are based on at least $10^4$ independent
instances of graph $D$ and are log-binned to the base $1.2$. Probabilities are
given according to the gray scale on the right of each panel, where darker is
closer to $1$.}
\label{fig:fig2}
\end{figure}

Interestingly, a similar situation occurs when $x_A(0)$ remains fixed, along
with $p$ and $r$, while $\lambda$ and $\mu$ are made to vary relative to each
other. This is shown in Fig.~\ref{fig:fig2}, in whose panels we have $p=r=0.1$
and $x_A(0)=0.1$ but $\lambda$ and $\mu$ are given values one order of
magnitude apart: $\lambda=0.01$ and $\mu=10\lambda$ in panel (a), $\lambda=0.1$
and $\mu=\lambda/10$ in panel (b). Thus, the ratio $\lambda/\mu$ increases by
two orders of magnitude from one scenario to the next. In the first of these
scenarios it is uncertain whether the quasispecies will be led to extinction,
whereas in the second it almost certainly survives. The ratio $\lambda/\mu$
indicates how much more responsive the immune system is in reorganizing itself
to respond to the threat of the genotypes than it is in actually destroying
genotypes. What we see in the figure is that a heavy imbalance toward the former
almost certainly leads to dominance by the attacking quasispecies.

\begin{figure}
\vspace{0.20in}
\includegraphics[scale=0.35]{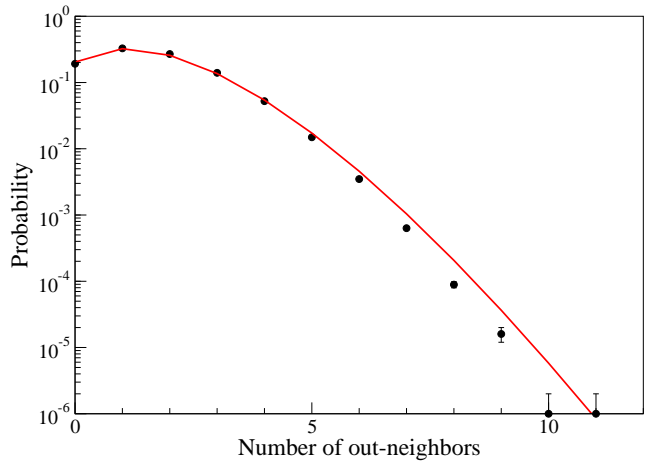}
\caption{(Color online) Probability distribution of a genotype's number of
out-neighbors in set $A$, ignoring self-loops. Data are given for $p=0.1$ and
are based on $10^4$ independent instances of graph $D$. As defined, the
probability that an edge exists between two genotypes in $D$ depends on the
Hamming distance between them. If this probability were independent of the two
genotypes, the resulting distribution would be the Poisson distribution, here
shown as a solid line for the same mean of $1.59$ that is given for $p=0.1$ by
Eq.~(\ref{eq:outAA}). If $z$ is the expected number of out-neighbors given by
this equation, then the Poisson distribution of mean $z$ requires that edges
exist with the fixed probability $z/(2^L-1)=[(1+p)^L-1]/(2^L-1)$, which
approaches the probabilities we use only as $p$ nears $1$.}
\label{fig:fig3}
\end{figure}

Throughout the transition depicted in Figs.~\ref{fig:fig1} and~\ref{fig:fig2},
the bifurcation that the genotype population undergoes at about $t=2$ clearly
depends on the particular instance of graph $D$ being used. We study the role of
network topology on the quasispecies by initially continuing to focus on the
$p=r=0.1$ setting that is common to both Figs.~\ref{fig:fig1}
and~\ref{fig:fig2}. In this setting, the stationary-state distribution of a
genotype's number of out-neighbors is the one shown in Fig.~\ref{fig:fig3},
where only out-neighbors in set $A$ are counted (i.e., out-neighbors that are
genotypes as well, so only the value of $p$ matters) and self-loops are ignored
(so that $0$ is a possibility). This distribution is clearly concentrated on the
lowest numbers of out-neighbors and in this range it seems to be possible to
approximate it by the Poisson distribution of mean given by
Eq.~(\ref{eq:outAA}). However, the Poisson distribution arises only when edges
exist with the same probability for all node pairs (as in the original
Erd\H{o}s-R\'{e}nyi model \cite{er59} and its directed variant \cite{k90}), and
in fact we see in the figure that the two distributions differ markedly for the
higher numbers of out-neighbors.

The distribution shown in Fig.~\ref{fig:fig3} indicates that a randomly
chosen genotype is likely to be able to mutate only into a small number of other
genotypes. If idiotypes were altogether absent this would entail excellent
survival chances for the quasispecies, meaning in particular that the wild type
would be able to climb from its initial relative abundance of
$1/2^L\approx 10^{-3}$ toward a stationary-state relative abundance of the order
of $10^{-1}$ \cite{bds12}. In the presence of idiotypes, however, one must look
at how network topology affects the quasispecies from a closer perspective,
paying special attention to how the wild type's number of out-neighbors affects
its stationary-state relative abundance, $x_1$.

\begin{figure}
\vspace{0.20in}
\includegraphics[scale=0.35]{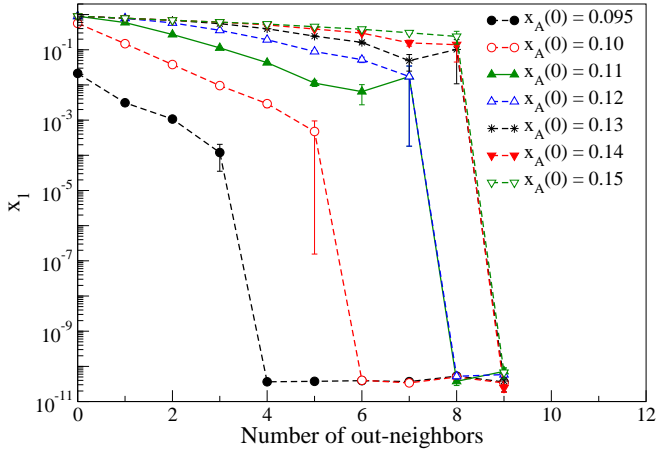}
\caption{(Color online) Stationary-state relative abundance of the wild type
($x_1$) as a function of its number of out-neighbors in set $A$, ignoring the
self-loop. Data are given for $p=r=0.1$ and $\lambda=\mu=0.1$, and are based on
$10^5$ independent instances of graph $D$.}
\label{fig:fig4}
\end{figure}

\begin{figure}
\vspace{0.25in}
\includegraphics[scale=0.35]{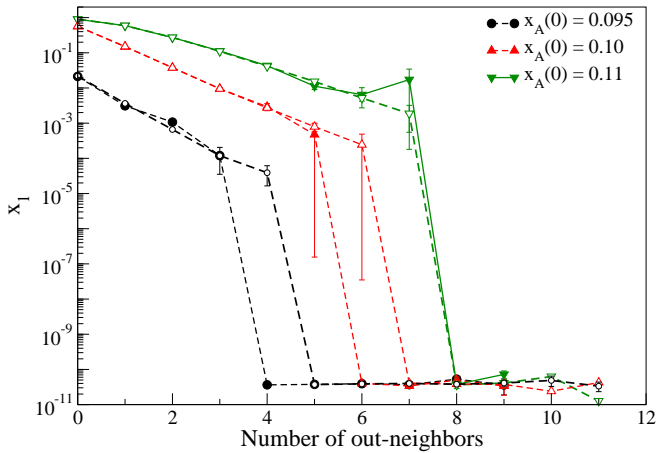}
\caption{(Color online) Stationary-state relative abundance of the wild type
($x_1$) as a function of its number of out-neighbors in set $A$, ignoring the
self-loop. Data are given for $p=r=0.1$ and $\lambda=\mu=0.1$, and are based on
either $10^5$ (filled symbols) or $10^6$ (empty symbols) independent instances
of graph $D$.}
\label{fig:fig5}
\end{figure}

Figure~\ref{fig:fig4} shows the average value of $x_1$ as a function of the
wild type's number of out-neighbors in set $A$ and ignoring the self-loop. All
data in this figure are still relative to $p=r=0.1$ and $\lambda=\mu=0.1$. Each
plot corresponds to a different value of the initial genotype population,
beginning at $x_A(0)=0.095$ and proceeding through $x_A(0)=0.15$ (i.e., up to
inside the transition qualitatively depicted in Fig.~\ref{fig:fig1}). In those
instances of graph $D$ in which the wild type has no out-neighbors, the value of
$x_1$ is influenced from outside exclusively by mutations into the wild type and
by the action of the idiotypes. Under these circumstances, it is clear from the
figure that the wild type recovers from its initial situation of uniform
dilution with respect to the other genotypes, i.e., from an initial relative
abundance of $x_A(0)/2^L\approx 10^{-4}$. Instances with larger numbers of
out-neighbors at the wild type, on the other hand, reflect the effect of
mutations of the wild type into other genotypes as well, leading to
progressively smaller values of $x_1$ until a precipitous drop to some value
between $\delta/10$ and $\delta$ is reached (cf.\ Section~\ref{sec:results}),
indicating the total absence of $D$ instances in which the wild type has any of
the corresponding numbers out-neighbors. Some of this figure's plots are
repeated in Fig.~\ref{fig:fig5} alongside new plots for the same values of
$x_A(0)$ but ten times as many instances of graph $D$, which postpones the
aforementioned drop as $D$ instances appear in which the wild type has a larger
number of out-neighbors. Taken together, the two figures seem to support the
presence of an exponential decay of $x_1$ with the wild type's number of
out-neighbors. However, as already suggested by the data in Fig.~\ref{fig:fig1},
such decay is ever slower as $x_A(0)$ increases, so the wild type is ever more
resilient to the possibility of mutation into a larger number of genotypes.

\begin{figure}
\vspace{0.20in}
\includegraphics[scale=0.35]{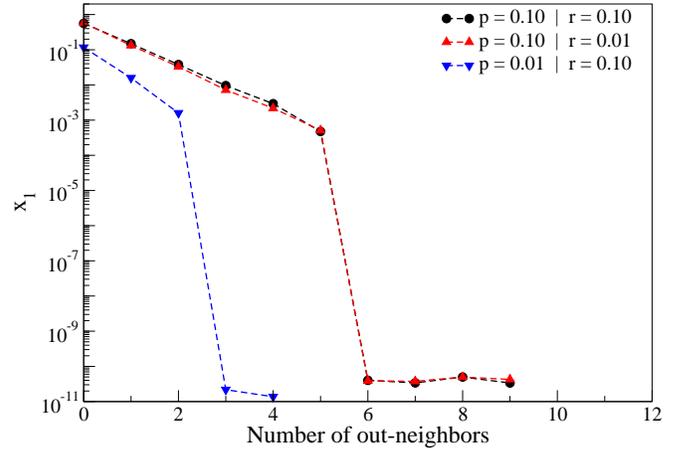}
\caption{(Color online) Stationary-state relative abundance of the wild type
($x_1$) as a function of its number of out-neighbors in set $A$, ignoring the
self-loop. Data are given for $\lambda=\mu=0.1$ and $x_A(0)=0.1$, and are based
on $10^5$ independent instances of graph $D$ when $p=0.1$, $10^6$ instances
when $p=0.01$.}
\label{fig:fig6}
\end{figure}

Varying the value of $p$ and $r$ relative to each other can provide further
insight into wild-type survival, since these are the parameters regulating
edge density and also mutation or stimulation on those edges that do exist in
a particular instance of graph $D$. We give data for this in
Fig.~\ref{fig:fig6}, which continues to be relative to $\lambda=\mu=0.1$ but now
focuses exclusively on the $x_A(0)=0.1$ case of Figs.~\ref{fig:fig4}
and~\ref{fig:fig5}. Note, first, that using $p=0.1$ with $r=0.01$ has no
relevant effect with respect to the $p=r=0.1$ case. The reasons for this can be
grasped from Eqs.~(\ref{eq:outAB}) and~(\ref{eq:outBB}): According to these
equations, decreasing $r$ from $0.1$ to $0.01$, though entailing a difference by
one order of magnitude in the value of this probability, impacts the expected
number of out-neighbors a genotype or an idiotype has inside set $B$ much more
modestly. In fact, these variations are in both cases from about $2.59$ to about
$1.01$ and seem to affect genotype-idiotype interaction very little.

On the other hand, using $p=0.01$ with $r=0.1$, while affecting only the
quasispecies side of the network in comparison to the $p=r=0.1$ case, leads to
very different results. By Eq.~(\ref{eq:outAA}), now a genotype's expected
number of out-neighbors in set $A$ drops from $1.59$ to $0.1$, therefore by one
order of magnitude as well. Genotypes are then very sparsely interconnected to
one another, but in spite of this we see in Fig.~\ref{fig:fig6} that the wild
type manages to survive for those instances of graph $D$ in which it has a very
small number of out-neighbors. The pattern of survival is significantly
different from the previous one, which may at first seem as a surprise given
that the pull of idiotypes on genotypes continues to be the same and all that
has changed is the density in genotype interconnection and the probabilities
that mutations occur. However, the resulting expected sparsity does not rule out
those relatively rare $D$ instances in which the wild type is fed by mutations
from at least one other genotype. Combined with the very low number of genotypes
into which the wild type itself may mutate, this unexpected existence of edges
incoming to the wild type ensures survival.

It follows from these analyses of the data in Fig.~\ref{fig:fig6} that, as in
the case of the isolated quasispecies network \cite{bds12}, probability $p$ is
instrumental in determining the fate of the wild type. Contrasting with that
case, however, the wild type is no longer guaranteed to survive for arbitrarily
low values of $p$. Instead, it seems that some minimum value is required for the
quasispecies to resist the action of the idiotypes, i.e., for the wild type to
escape being diluted into the remainder of the quasispecies.

\section{Conclusion}
\label{sec:concl}

Quasispecies studies have invariably concentrated on characterizing two distinct
regimes for the evolution of genotypes through time, one in which the wild type
survives in the quasispecies, another in which the wild type becomes diluted and
no more abundant than any of the quasispecies' other genotypes. Distinguishing
between the two regimes has been a matter of selecting the right perspective
from which to model replication errors, or mutations. In our own previous model,
for example, the single parameter $p$ can be used to characterize quasispecies
survival or demise: Increasing $p$ progressively leads the wild type to a
situation in which its relative abundance cannot rise above those of the other,
less fit genotypes \cite{bds12}. This focus on the two extremal regimes of
survival and demise has been motivated by the theory's purported use in the
modeling of viral populations, along with its appeal as a potential aid in the
discovery of therapies \cite{d09,la10}.

However, in our view the picture has clearly been incomplete, particularly when
viral pathogens are the motivation, because the quasispecies' interaction with
the host's immune system seems to have been left completely aside. The present
work constitutes an attempt to take the immune system into account and study its
effect on quasispecies behavior. Our model uses the same network as \cite{bds12}
to represent the interacting genotypes and interconnects it to another network
built on top of the so-called idiotypes, some of the fundamental motifs of the
host's immune response. Like the quasispecies network, this new, idiotypic
network is self-adapting (based on the idea of complementarity among idiotypes)
and potentially effective in destroying genotypes (based on the idea that
genotypes and idiotypes are mutually complementary to some degree).

The resulting network is a continuous-time dynamical system whose variables,
representing the relative abundances of genotypes and idiotypes, are coupled
with one another according to a random-graph model. The system's behavior
depends on four parameters (the probabilities $p$ and $r$ and the rates
$\lambda$ and $\mu$), as well as on the initial relative abundance of genotypes,
$x_A(0)$, which together give rise to a phase space of vast proportions. In view
of this, we have concentrated on analyzing a specific niche inside which both
survival and destruction of the quasispecies can be examined side by side.

Our exploration of this particular niche has highlighted the existence of two
main factors influencing quasispecies survival. The first one is $x_A(0)$
itself, or equivalently the interplay between $\lambda$ and $\mu$. If genotypes
are reasonably abundant in the beginning with respect to idiotypes, or if
idiotypes eliminate genotypes much more slowly than the idiotypes adapt their
abundances to meet the challenge of the mutating genotypes (i.e.,
$\mu\ll\lambda$), then it is possible for the quasispecies to survive. The
second factor is the topology of the quasispecies network, governed by the $p$
parameter. Unlike the case of the isolated quasispecies network, in which very
low values of $p$ are practically a guarantee that the wild type will nearly
dominate the quasispecies even from an initial situation of total dilution, the
presence of the idiotypic network makes things quite different. Specifically, a
non-negligible value of $p$ seems necessary to ensure that the quasispecies
network is sufficiently dense, and mutations sufficiently likely, for genotypes
to change into one another and help the wild type evade the action of the
idiotypic network.

Further research will concentrate on revisiting Eqs.~(\ref{eq:XB})
and~(\ref{eq:XA2}) with the goal of modeling the action on the immune system of
those quasispecies, such as that of HIV, that work toward depleting the system's
supply of idiotypes. While the required changes may turn out to be conceptually
simple, it seems unavoidable that at least one new rate parameter will be
needed, thus enlarging the model's phase space even further. Finding an
appropriate parameter niche to explore, however, may provide interesting insight
into how genotype mutation, idiotype stimulation, and idiotype destruction
affect one another.

\begin{acknowledgments}
We acknowledge partial support from CNPq, CAPES, a FAPERJ BBP grant, and the
joint PRONEX initiative of CNPq/FAPERJ under contract No.~26-111.443/2010.
\end{acknowledgments}

\bibliography{inet}
\bibliographystyle{apsrev}

\end{document}